# Dependence versus Conditional Dependence in Local Causal Discovery from Gene Expression Data

Eric V. Strobl[1,*] and Shyam Visweswaran[1]

[1]*Department of Biomedical Informatics, University of Pittsburgh*
*5607 Baum Boulevard, Pittsburgh, PA 15206, USA*

**ABSTRACT**
**Motivation:** Algorithms that discover variables which are causally related to a target may inform the design of experiments. With observational gene expression data, many methods discover causal variables by measuring each variable's degree of statistical dependence with the target using dependence measures (DMs). However, other methods measure each variable's ability to explain the statistical dependence between the target and the remaining variables in the data using conditional dependence measures (CDMs), since this strategy is guaranteed to find the target's direct causes, direct effects, and direct causes of the direct effects in the infinite sample limit. In this paper, we design a new algorithm in order to systematically compare the relative abilities of DMs and CDMs in discovering causal variables from gene expression data.
**Results:** The proposed algorithm using a CDM is sample efficient, since it consistently outperforms other state-of-the-art local causal discovery algorithms when samples sizes are small. However, the proposed algorithm using a CDM outperforms the proposed algorithm using a DM only when sample sizes are above several hundred. These results suggest that accurate causal discovery from gene expression data using current CDM-based algorithms requires datasets with at least several hundred samples.
**Availability:** The proposed algorithm is freely available at https://github.com/ericstrobl/DvCD.

## 1 INTRODUCTION

Causality refers to a relation between a set of variables and another set of variables such that the latter is understood to be a consequence of the former (Bunge, 1959). The concept of causality plays a central role in the sciences and has traditionally been determined by randomized experiments, where experimental units are randomly assigned to different groups. However, these experiments can sometimes be practically impossible or unethical to perform. For example, an investigator interested in discovering the causes of schizophrenia cannot manipulate the genetics or long-term environmental conditions of human subjects. Moreover, experimental results obtained from animal studies may not be directly applicable to humans.

The shortcomings of randomized experiments have led investigators to develop approaches that can supplement experimentation. In the context of biomedicine, some have noted the abundance of observational as opposed to experimental data. For example, publicly available gene expression datasets such as TCGA (Weinstein, et al., 2013) and GEO (Edgar, et al., 2002) have been steadily growing larger in sample size and more comprehensive in the genes they measure. Investigators have thus explored the use of readily available observational gene expression data to discover causal genes, where the term "causal" refers to those variables that either change the target (e.g., a diagnosis, expression of another gene) or its effects when experimentally manipulated (Bunge, 1959). The earliest algorithms that were developed to identify casual variables in observational data used dependence measures (DMs). For example, algorithms identified casual genes by finding those with significant univariate associations to the target using the t-statistic (Tusher, et al., 2001) or Fisher-score (Hastie, et al., 2001) on the assumption that causal genes are likely to be a subset of the genes which display statistical dependence with the target. Later methods such as Relief-F (Kira and Rendell, 1992) and BAHSIC (Song, et al., 2007) identified potential casual variables that have a multivariate effect on the target. No single strategy consistently outperformed all others in every expression dataset. However, the multivariate methods were able to discover variables that univariate methods failed to identify (Zucknick, et al., 2008).

In more recent research, investigators have developed algorithms that employ conditional dependence measures (CDMs) for discovering causal variables. In contrast to DM-based algorithms, CDM-based algorithms are based on directed graphical models and have certain causal guarantees (details in section 2.1) (Spirtes, et al., 2000). While the earliest methods such as IAMB (Tsamardinos and Aliferis, 2003) and PC (Spirtes, et al., 2000) were sample inefficient and/or did not scale up to high-dimensional data, newer methods have overcome these shortcomings. For example, HITON-MB discovers the Markov blanket (see Section 2.1) of a target by employing univariate dependence tests before performing conditional dependence tests (Aliferis, et al., 2003). Other algorithms such as kernel-based methods have been developed to rank potential variables in the Markov blanket in order to help prioritize experimentation (Strobl and Visweswaran, 2013).

The hypothesized superior ability of CDMs in detecting causal relationships compared to DMs has unfortunately not been clear in gene expression data despite the algorithmic advances for several reasons. First, the structures of algorithms using either DMs or CDMs are different which makes comparison of the relative capabilities of dependency and conditional dependency difficult. Most investigators have thus restricted themselves to comparing DM-based algorithms with other DM-based algorithms or CDM-based algorithms with other CDM-based algorithms. Second, many CDM-based algorithms require the variables to be discretized (Statnikov, et al., 2010) while most DM-based algorithms do not. Thus differences in performance may be confounded by information loss due to discretization. Third, gold standards of causal relationships from real gene expression data have not been available until recently

---

(Statnikov, et al., 2012) which has forced many investigators to draw conclusions from synthetic datasets that may not capture important relations which exist in real data. Accurately determining performance differences between DM-based and CDM-based algorithms in gene expression data have to address these three problems.

In this paper, we compare DM-based and CDM-based algorithms by (1) developing a standardized algorithm that can use either a DM or a CDM, where the only difference between the two measures is a matrix inversion, (2) using the universal RBF kernel in DM and CDM in order to detect non-linear relationships from both discrete and continuous data in a fully multivariate fashion, and (3) evaluating the algorithms in their ability to detect true Markov blankets using real gene expression data. Our results indicate that the algorithm using the DM significantly outperforms the CDM counterpart for small sample sizes, but the algorithm using the CDM outperforms for large sample sizes. These results suggest that both DMs and CDMs are useful in detecting causal relationships from gene expression data, but the performance of the specific measure is heavily influenced by the sample size.

## 2 METHODS

### 2.1 Overview

From here on, upper-case letters in italics will denote single variables, and upper-case letters in bold italics will denote sets of variables. The Markov blanket ($MB(Y)$) consists of the direct causes, direct effects, and direct causes of the direct effects of the target variable (see Figure 1). It also represents the smallest subset of variables that can maximally explain the target variable in the dataset (Pearl, 1988). We wish to compare the *intrinsic* abilities of DMs and CDMs in detecting $MB(Y)$ in real gene expression data independent of any differences in algorithmic structure. In order to do so, we develop a single algorithm that can employ either a DM or a CDM and scale to high-dimensional data. The ability of the algorithm in detecting $MB(Y)$ when using either the DM or CDM is then evaluated on 6 synthetic datasets and 24 real gene expression datasets.

The algorithm takes advantage of the kernel method to efficiently compute DMs or CDMs in high dimensional space. It works by constructing a kernel of the target variable and then another kernel of the predictors. The importance of each predictor is determined by removing the predictor from the second kernel (i.e., creating a "predictor exclusion kernel") and then measuring either the amount of dependence or conditional dependence between the two kernels. The tested predictor is then placed back into the second kernel, and the process is repeated by removing a different single predictor until all predictors have been tested.

The DM we use in this paper is the cross-covariance operator in kernel induced space. Cross-covariance refers to the linear covariance between two sets of random variables, and kernels are symmetric, positive semidefinite matrices that represent the inner product of potentially high dimensional representations of the original data (Hofmann, et al., 2008). The DM can measure non-linear dependencies between the target and predictors by projecting each set of variables into two high dimensional spaces using kernels and then measuring the degree of linear cross-covariance between the two spaces (Gretton, et al., 2005; Song, et al., 2012). Practically, the DM is computed by taking the trace of the inner product of the target kernel and predictor exclusion kernel. Note that removing a variable in $MB(Y)$ from the predictor exclusion kernel may decrease the value of the DM.

The CDM in this paper is the conditional cross-covariance operator in kernel-induced space. The conditional cross-covariance refers to the linear covariance between two sets of random variables given a third set of random variables. When specifically measuring the conditional cross covariance between the target and the target given the predictors, the CDM computes the residual errors after regressing the target on the predictors (Fukumizu, et al., 2009). Non-linear kernels are introduced to measure non-linear conditional dependencies, and the measure is practically computed by taking the trace of the inner product of the target kernel and the *inverted* predictor exclusion kernel. Thus, the only difference between the DM and CDM is a matrix inversion. Note that removing a variable in $MB(Y)$ from the predictor exclusion kernel may increase the residual errors of regression and thus the value of the CDM. We next provide details of the algorithm.

### 2.2 Background

A causal directed graph is a graph where nodes represent variables and directed edges denote causal relationships between pairs of variables (Friedman and Koller, 2009). We define the joint probability distribution of the directed graph as a globally normalized product of non-negative interventional potential functions such that the discrete case is given by:

$$p(\mathbf{Z}) = \frac{\prod_{i=1}^{d} \phi(Z_i|pa(Z_i))}{\sum_{\mathbf{Z}} \prod_{i=1}^{d} \phi(\mathbf{Z}|pa(Z_i))},$$

where $\mathbf{Z}$ refers to the entire dataset of $d$ variables, $Z_i$ to one variable in $\mathbf{Z}$ with column index $i$, and $\phi(\mathbf{Z}|pa(Z_i))$ to the non-negative interventional potential assigned to $Z_i$ and its parents $pa(Z_i)$. The denominator ensures that the sum over all possible configurations of $\mathbf{Z}$ is unity. Moreover, the continuous case follows by replacing the summation with integration and the specific values of $Z_i$ by intervals. Further details of this definition can be found in Schmidt and Murphy (2009). Note that we obtain a directed acyclic graph (DAG) in the special case when the directed edges are constrained such that there are no cycles in the graph. In this case, the interventional potentials represent conditional probabilities, since they satisfy a local normalization constraint:

$$\forall pa(Z_i), \sum_{Z_i} \phi(Z_i|pa(Z_i)) = 1.$$

We however do not restrict ourselves to the DAG in this paper, since many biological pathways are known to be cyclic.

The Markov blanket of a variable $Y$ is defined as its direct causes, direct effects, and direct causes of the direct effects (Figure 1). The variable $Y$ in a directed graph is conditionally independent of all the other $d-1$ variables represented as $\mathbf{X}$ given its Markov blanket $MB(Y)$:

$$Y \perp\!\!\!\perp \{\mathbf{X} \setminus MB(Y)\}|MB(Y) \Leftrightarrow Y \perp\!\!\!\perp \mathbf{X}|MB(Y).$$

The proof of the equivalence is given as a proposition in the Supplementary Materials. For clarity, we reiterate in mathematical notation that $\{\mathbf{X} \cap Y\} = \emptyset$ and $\{\mathbf{X} \cup Y\} = \mathbf{Z}$.

Discovering the minimal subset of variables in $\mathbf{X}$ that render $Y$ and $\mathbf{X}$ independent may provide a more principled approach to uncover causal biological relationships than discovering the variables in $\mathbf{X}$ that are most dependent on $Y$ for several reasons. First, assuming the direct effects, direct causes, and direct causes of the direct effects are measured in the dataset, and the variables in $MB(Y)$ are correctly discovered, specific manipulations of the direct causes in $MB(Y)$ are guaranteed to change the interventional potential of the target (Pearl, 2009). Specific manipulations of the direct causes or direct effects of the direct causes are also guaranteed to change the interventional potentials of the target's direct effects. Moreover, while the direct effects and direct causes must be dependent on the target in the infinite sample limit, other variables that are not in $MB(Y)$ can also be dependent on the target (Guyon, et al., 2007). Manipulations of those variables that are most dependent on the target are thus not guaranteed to change the interventional potential of the target or any of its direct consequences in contrast to manipulations of those variables in $MB(Y)$.

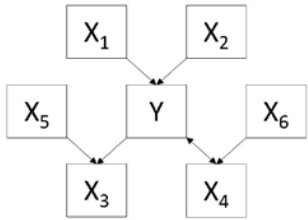

**Fig. 1.** An example of a Markov blanket. Variable $Y$ is the target. Variables $X_1$ and $X_2$ are direct causes of Y, $X_3$ is a direct effect of Y, $X_4$ is both a direct cause and direct effect, and $X_5$ and $X_6$ are the direct effects of the direct causes.

One way of measuring conditional dependence between arbitrary interventional potentials or distributions involves computing the trace of the empirical conditional cross-covariance operator between $X$ and $Y$ within reproducing kernel Hilbert spaces (RKHSs) (Fukumizu, et al., 2009). Specifically, we can map $X$ and $Y$ into RKHSs $\mathcal{F}$ and $\mathcal{G}$ respectively using two positive semidefinite kernels $K_X: \mathcal{X} \times \mathcal{X} \to \mathbb{R}$ and $K_Y: \mathcal{Y} \times \mathcal{Y} \to \mathbb{R}$. There then exists a conditional cross-covariance operator $\Sigma_{YY|X}: \mathcal{G} \to \mathcal{G}$ for any function $g \in \mathcal{G}$ such that:

$$\langle g, \Sigma_{YY|X} g \rangle_\mathcal{G} = \mathbb{E}_X \left[ Var_{Y|X}[g(Y)|X] \right],$$

which represents the residual errors of predicting $g(Y)$ with $X$.

We now denote $X_S$ as some subset of the variables in $X$. Then, the conditional cross-covariance operator exhibits the following property: $\Sigma_{YY|X_S} \geq \Sigma_{YY|X}$, where the order is determined by the trace operator, and the equality holds when the subset $X_S$ includes $MB(Y)$ so that $Y \perp\!\!\!\perp X|X_S$.

Empirically, we can compute the kernel matrices $K_{X_S}$ and $K_Y$ from a sample size of $n$ drawn independently and identically distributed (i.i.d.) from $P(X, Y)$. The trace of the empirical conditional cross-covariance operator after dropping constants is then defined by:

$$M^1 = tr\big(G_Y(G_{X_S} + n\varepsilon I_n)^{-1}\big), \quad (1)$$

where $G_{X_S} = \left(I_n - \frac{1}{n}1_n 1_n^T\right) K_{X_S} \left(I_n - \frac{1}{n}1_n 1_n^T\right)$ where $n$ is the sample size, $I_n$ is an $n \times n$ identity matrix, and $1_n$ is a vector of ones. The regularization term $\varepsilon \to 0$ is added for the inversion.

Dependence can be measured in a similar fashion (Gretton, et al., 2005; Song, et al., 2012), whereby the cross-covariance operator $\Sigma_{YX}: \mathcal{F} \to \mathcal{G}$ is defined for any functions $f \in \mathcal{F}$ and $g \in \mathcal{G}$ so that:

$$\langle g, \Sigma_{YX} f \rangle_\mathcal{G} = \mathbb{E}_{XY}[(f(X) - \mathbb{E}_X[f(X)])(g(Y) - \mathbb{E}_Y[g(Y)])].$$

The cross-covariance operator then exhibits a property opposite to that of the conditional cross-covariance operator, namely: $\Sigma_{YX_S} \leq \Sigma_{YX}$, where the order is again determined by the trace operator, and the equality holds when $X_S$ and $Y$ are as dependent as $X$ and $Y$. The trace of the empirical conditional cross-covariance operator is then defined by:

$$M^2 = tr\left(G_Y(G_{X_S} + n\varepsilon I_n)\right) = tr(G_Y G_{X_S}) + n\varepsilon tr(G_Y) \\ \propto tr(G_Y G_{X_S}), \quad (2)$$

where the term $n\varepsilon tr(G_Y)$ is a constant and thus can be dropped.

Importantly, the only difference between computing the trace of the empirical conditional cross-covariance operator and the trace of the empirical cross-covariance operator in RKHSs is a matrix inversion. This property allows us to accurately assess any performance differences between the two measures. In contrast, many other methods of measuring dependence and conditional dependence are less similar to each other, since measuring dependence in most methods requires the comparison of two groups of variables $Y$ and $X_S$, while measuring conditional dependence requires the comparison of three groups $\{X\setminus X_S\}$, $Y$ and $X_S$ (Bergsma, 2004). Without loss of generality, the empirical measures in the proposed method only require the comparison of $Y$ and $X_S$ in both cases.

## 2.3 Algorithms

### 2.3.1 Method using Predictor Exclusion Kernels

We now use the term kernel conditional dependence measure (K-CDM) to denote Equation 1 and kernel dependence measure (K-DM) to denote Equation 2. The symbols $M^*(Y, X_S, \sigma)$ represent $M^1$ or $M^2$ evaluated with $Y$, $X_S$, and $\sigma$ such that $\sigma$ is the set of kernel hyperparameters (if any). We propose to discover the Markov blanket by iteratively computing the K-CDM or K-DM value between $Y$ and $\{X\setminus X_i\}$ which denotes a specific variable $X_i$ removed from the dataset $X$ (Algorithm 1). As a result, $K_{X_S}$ in Equation 1 and 2 is specifically $K_{\{X\setminus X_i\}}$ and is defined as a "predictor exclusion kernel" in each iteration.

The algorithm works as follows. For every variable, it computes a K-CDM or K-DM value by generating a predictor exclusion kernel $K_{\{X\setminus X_i\}}$ and then stores the value in $V$. Subsequently, the variables in $X$ are sorted according

---

**Algorithm 1: Method using Predictor Exclusion Kernels**
1. **Input:** Target feature $Y$, non-target features $X$
2. **Output:** Non-target features in descending order $X^\dagger$
3. $V \leftarrow \emptyset$
4. **for** $i = 1: d-1$
5. $\quad v \leftarrow M^*(Y, \{X\setminus X_i\}, \sigma), \sigma \in \Xi$
6. $\quad V \leftarrow V \cup v$
7. **end**
8. $X^\dagger \leftarrow sort(V, X)$

---

to the values in $V$ such that variables associated with larger K-CDM or lower K-DM values when removed are ranked lower than those variables with lower K-CDM or higher K-DM values when removed. The algorithm thus gives a lower rank to variables that can better explain the dependence between $Y$ and $X$ in the K-CDM case, or gives a lower rank to variables that exhibit a stronger degree of dependence to $Y$ in the K-DM case.

Algorithm 1 has several noteworthy properties. First, we can prove that the lowest ranked variables in $X^\dagger$ are guaranteed to contain the variables in $MB(Y)$ in the infinite sample limit when K-CDM is used.

**Theorem 1 (Correctness).** Assume that (1) dataset $X$ contains $MB(Y)$, (2) K-CDM is defined by Equation 1 such that both $K_{X_S}$ and $K_Y$ are positive semidefinite kernels and $K_{X_S}$ satisfies the universal approximating property, and (3) $\{X \cup Y\}$ has an infinite sample size, where the samples are drawn i.i.d. from a probability distribution faithful to a directed cyclic or acyclic graph. Then, the lowest ranked variables in $X^\dagger$ from Algorithm 1 will include $MB(Y)$.

**Proof.** First, a lower value returned from Equation 1 denotes a smaller amount of residual errors when predicting $g(Y)$ with $X$ than a higher value. Second, $MB(Y)$ is maximally predictive of $g(Y)$ by the definition of a Markov blanket, since $Y$ is conditionally independent of $X$ given $MB(Y)$. As a result, K-CDM is guaranteed to return a higher value every time a variable in $MB(Y)$ is tested for removal in line 5 compared to a variable *not* in $MB(Y)$ if and only if (1) $X$ contains variables in and not in $MB(Y)$, (2) K-CDM is defined by Equation 1 such that both $K_{X_S}$ and $K_Y$ are positive semidefinite kernels and $K_{X_S}$ satisfies the universal approximating property, and (3) an infinite sample of data points are drawn i.i.d. from a probability distribution faithful to the causal directed cyclic or acyclic graph. In line 8, variables associated with smaller K-CDM values will then be ranked lower in $X^\dagger$. In the special case where $MB(Y) = X$, all the variables in $MB(Y)$ will be in $X^\dagger$ regardless of the K-CDM values, since all the variables in $X$ will be placed into $X^\dagger$ in line 8.□

Also, Algorithm 1 is similar to performing one pass of backward elimination, where all variables are eliminated after each feature is tested for removal in $K_{\{X\setminus X_i\}}$ (Strobl and Visweswaran, 2013). However, unlike backward elimination, the variables are sorted according to their K-CDM or K-DM values in the end instead of being eliminated indiscriminately. The proposed method thus has a smaller time complexity than performing multiple passes of backward elimination (full complexity analysis is given in section 2.3.4) and is scalable to large gene expression datasets containing tens of thousands of variables.

### 2.3.2 Comparison to other Markov Blanket Discovery Algorithms

Many algorithms have been proposed to discover $MB(Y)$. However, these algorithms share common themes which carry weaknesses. First, the data efficient algorithms such as MMMB and HITON-MB utilize a two stage process, where they find the direct effects and direct causes of a node, and then the direct effects of the direct causes (Aliferis, et al., 2010). These two stages can increase the probability of including an incorrect variable in $MB(Y)$, since an error is more likely to occur in the second stage when an error is made in the first stage. Second, the data efficient algorithms require a parameter that specifies the size of the conditioning set which carries an exponential time complexity as the size of this set is increased (Statnikov, et al., 2013). This requirement can prevent the algorithm from discovering some multivariate relationships, since certain variables may only be conditionally dependent on the target when given a very large conditioning set. Third, the scalable tests of conditional dependence used in these methods in practice such as the G² test require discretization and/or the comparison of three groups of variables $\{X\backslash X_S\}$, $Y$ and $X_S$ (Statnikov, et al., 2010). Unfortunately, the former results in information loss and the latter can increase the Type I and Type II error rates with a finite sample size.

*2.3.3 Method using Predictor Inclusion Kernels*

We were also interested in comparing how well Algorithm 1 with K-CDM performs against a univariate K-DM. To this end, we propose Algorithm 2 which is similar to Algorithm 1 except in line 5, where $M^2(\cdot)$ is restricted to K-DM using "predictor inclusion kernels" such that only a single variable $X_i$ is used to construct $K_{X_i}$ in each iteration. We do not use K-CDM in line 5 because Algorithm 2 with K-CDM is not guaranteed to contain $MB(Y)$ in the infinite sample limit, as it cannot detect any multivariate relationships between the variables in $MB(Y)$. Note that variables associated with larger K-DM values when removed are ranked lower in this case, as opposed to higher when using Algorithm 1 with K-DM.

---

**Algorithm 2: Method using Predictor Inclusion Kernels**

1. **Input:** Target $Y$, predictors $X$
2. **Output:** Predictors in descending order $X^\dagger$
3. $V \leftarrow \emptyset$
4. **for** $i = 1: d - 1$
5. $\quad v \leftarrow M^2(Y, X_i, \sigma), \sigma \in \Xi$
6. $\quad V \leftarrow V \cup v$
7. **end**
8. $X^\dagger \leftarrow sort(V, X)$

---

*2.2.4 Time and Memory Complexities*

Algorithm 1 with K-CDM has time complexity $O(dn^3)$ where $d$ represents the total number of variables and $n^3$ represents the inversion of the kernel $G_{\{X\backslash X_i\}}$ when calculating K-CDM. On the other hand, Algorithm 1 and 2 with K-DM both have time complexities $O(dn^2)$, since no matrix inversion is required with $G_{\{X\backslash X_i\}}$ or $G_{X_i}$. If we instead decide to remove $1 - \beta$ of $X_S$ at every iteration for a total of $k$ iterations in a backward elimination scheme, the method takes $(k\beta^{k-1}dn^3)$ for K-CDM and $O(k\beta^{k-1}dn^2)$ for K-DM. Algorithm 1 and 2 are thus more time efficient than backward elimination.

The memory complexity of both Algorithm 1 and 2 as well as backward elimination are $O(n^2)$ due to the storage of one $n \times n$ kernel $G_{\{X\backslash X_i\}}$ or $G_{X_i}$ at each iteration.

### 2.4 Datasets

We obtained 6 datasets of expert-designed models titled Child, Alarm, Insurance, Pigs, Link, and Gene. We downloaded the "s5000_v1" files from http://www.dsl-lab.org/supplements/mmhc_paper/mmhc_index.html, where each file contains 5000 samples (Tsamardinos, et al., 2006). The samples were generated by sampling i.i.d. from the positive, discrete joint probability distributions of the respective models. The Child, Insurance and Alarm datasets were considered low dimensional at 20, 27, and 37 variables, respectively. On the other hand, the Pigs, Link and Gene datasets were considered high dimensional at 441, 724 and 801 variables, respectively.

We also obtained 20 real gene expression datasets in order to evaluate the performance of the algorithms as feature selectors for a support vector machine (SVM). These datasets were obtained from a variety of sources but are all well-established datasets of cancer. A summary of the datasets is presented in Table S1 (see Supplementary Materials). Examples include the leukemia dataset from Golub, et al. (1999), and the breast cancer dataset from Sorlie, et al. (2001).

We also obtained 2 real gene expression datasets called YEAST and ECOLI in order to assess the ability of the algorithms in detecting experimentally verified transcription factor-gene relationships. Both datasets were downloaded from www.nyuinformatics.org/downloads/supplements/CausalOrientation. Details of the datasets are given in (Statnikov, et al., 2012). The ECOLI dataset (Faith, et al., 2008; Narendra, et al., 2011) contains 907 samples, 4297 variables, and 140 relationships between transcription-factors and their target genes (TF-gene relationships) as identified from RegulonDB (version 6.4), a manually curated database of regulatory interactions obtained mainly through literature searches (Salgado, et al., 2013). The TF-gene relationships are restricted to those with "strong evidence" as classified by the database and include the downstream genes of multiple different TFs. The YEAST dataset (Faith, et al., 2008; MacIsaac, et al., 2006) contains 530 samples, 5520 variables, and 115 TF-gene relationships. The relationships include the downstream genes of multiple TFs. The relationships were identified from genome-wide ChIP-on-chip binding data at an alpha level of 0.001 and conserved within two species of the Saccharomyces genus.

### 2.5 Evaluation of Rankings and Accuracy

The rankings obtained from Algorithms 1 and 2 were normalized to compare variables with different sized Markov blankets as follows. If a continuous set of correct variables in $MB(Y)$ were identified, then those variables were given the same rank. However, a break in the correct identification led to a higher rank. For example, if variables 2, 3 and 4 are in $MB(Y)$ while 1, 5, and 6 are not, then an output of 6,3,5,4,2,1 in ascending order is converted to the ranking 5,4,3,2,2,1. The algorithm which provides a lower mean rank of $MB(Y)$ is then judged to perform better. In the example, the mean rank is 2.666, since the ranks of the variables in $MB(Y)$ are 4,2,2.

We also used the following accuracy measure in order to compare Algorithms 1 and 2 with three conditional dependence-based feature subset selection methods including IAMB, HITON-MB and MMMB:

$$A(X_c^\dagger, MB(Y)) = \frac{|X_c^\dagger \cap MB(Y)|}{|X_c^\dagger \cup MB(Y)|} * 100,$$

where $X_c^\dagger$ is the subset output from the conditional dependence algorithms or, for Algorithm 1 and 2, $X_c^\dagger$ is $X^\dagger$ clipped to the size of $MB(Y)$. For example, if variables 2, 3 and 4 are in $MB(Y)$ while 1, 5, and 6 are not, then an output of 6,3,5,4,2,1 from Algorithm 1 or 2 is converted to 4,2,1. Also, $|X_c^\dagger \cap MB(Y)|$ is the cardinality of the intersection of the subset $X_c^\dagger$ and the known $MB(Y)$, and $|X_c^\dagger \cup MB(Y)|$ is the cardinality of the union. Note that score $A$ is equal to 100 when the algorithm outputs the exact $MB(Y)$. On the other hand, decreasing the cardinality of $X_c^\dagger$ by failing to identify parts of the $MB(Y)$ or increasing the cardinality of $X_c^\dagger$ by random guessing both lead to a decrease in $A$.

### 2.6 Statistical Testing

All statistical comparisons were performed at an alpha level of 0.05. Independent and identically distributed samples were drawn from both synthetic and real gene expression datasets under different conditions as detailed in the results section. Each experiment was conducted using a within-subjects design such that every algorithm saw the same group of samples at each repetition. Mean ranks and accuracies in section 3.2 were then analyzed by paired Wilcoxon signed rank test for the expert-design models, since the distributions of the mean ranks were not approximately normally distributed. The same procedure was performed with mean support vector machine classification accuracies in section 3.3. On the other hand, the mean ranks of the downstream effects for the transcription factors in section 3.4 were approximately normally distributed and thus analyzed by a paired t-test instead.

## 3 RESULTS

### 3.1 Overview

We focused on comparing three methods including Algorithm 1 using K-CDM (which we now call ExcCD) or K-DM (ExcD), and Algorithm 2 using K-DM (IncD). Comparing ExcCD and ExcD allowed us to assess the relative abilities of *multivariate* CDMs and DMs in detecting $MB(Y)$. At the same time, comparing ExcCD and IncD allowed us to test whether *univariate* DMs could outperform multivariate CDMs, since searching for univariate dependencies may be more beneficial in noisy, high dimensional data (Lai, et al., 2006). We did not assess the ability of Algorithm 2 employing univariate CDMs, since such a setup has no theoretical guarantees in detecting $MB(Y)$ in the infinite sample limit.

We assessed the performance of each algorithm using three types of datasets. The first type included synthetic datasets generated from expert designed models, where the true Markov blankets are known and can thus be compared against each algorithm's outputs. The second type consisted of real gene expression datasets, where none of the variables in $MB(Y)$ are known. However, we can evaluate whether the algorithm is detecting $MB(Y)$ by first using the proposed algorithms as feature selection methods, then training a model using the top features (in this case, an SVM), and finally assessing the model's prediction accuracy. Prediction accuracy should be high, if the algorithm is truly detecting $MB(Y)$. Note that this is an approximate method, since there may be other variable combinations which can result in similar prediction accuracies as all the variables in $MB(Y)$, when there are a finite number of samples. Lastly, the third type of dataset also consisted of real expression data, but the task was to discover known transcription-factor gene relationships instead of evaluating classification accuracy. Since genes must be the direct effects of the transcription factor, part of $MB(Y)$ is known and could be evaluated against each of the proposed algorithms' outputs.

### 3.2 Expert-Designed Models

We ran the three algorithms along with three other well-known Markov blanket discovery algorithms including HITON-MB, MMMB, and IAMB on six publicly-available expert-designed datasets from Tsamardinos, et al. (2006). The alpha value for HITON-MB, MMMB and IAMB was fixed to 0.05, and the optimal conditioning set size was determined by 5-fold cross-validation from {2,3,4,5}. Three of the datasets were considered low-dimensional including Child, Insurance and Alarm at 20, 27, and 37 total variables respectively. Three other datasets including Pigs, Link and Gene were considered high dimensional at 441, 724 and 801 variables, respectively. The outputs of the algorithms were compared by randomly sampling the datasets 20 times with replacement with sample sizes of 25, 50, 100, 200, 300, 400 and 500. We randomly chose to discover the Markov blankets of variables 11, 18, 13, 126, 432 and 467 for Child, Insurance, Alarm, Pigs, Link and Gene, respectively. The performance of the algorithms was compared using the statistics described in section 2.5.

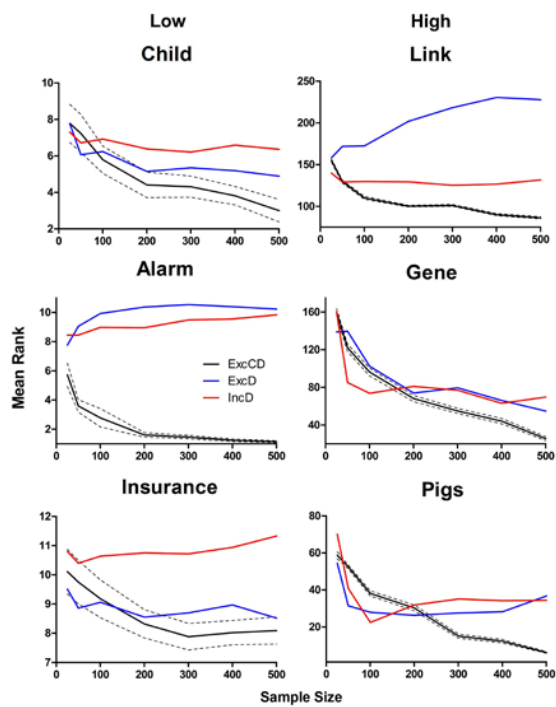

**Fig. 2.** Comparison of the ranks of the Markov blanket across ExcCD, ExcD and IncD. Solid lines represent mean rank (as defined in Section 2.5) and dotted lines represent the 95% confidence intervals. Note that ExcCD consistently outperforms ExcD and IncD when sample sizes are above 200.

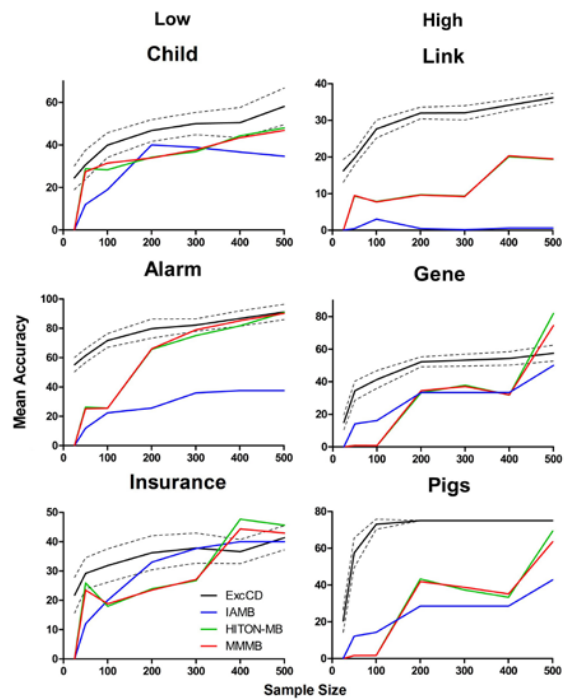

**Fig. 3.** Comparison of the accuracy in discovering the Markov blanket against other CDM-based methods. Solid lines represent mean accuracy (as defined in Section 2.5) and dotted lines represent the 95% confidence intervals. Note that ExcCD performs particularly well at low sample sizes (25-200), and the difference in accuracy is more pronounced in the high than low dimensional datasets.

Results are summarized in Figures 2 and 3. We found that ExcCD outperforms ExcD and IncD on all datasets when sample sizes are 400 and 500 as assessed by a paired Wilcoxon sign ranked test (all three pair-wise comparisons: W=21, p=0.031). The ExcCD method also significantly outperforms HITON-MB, MMMB and IAMB across all datasets with sample sizes of 25, 50, 100, 200 and 300 (W=21, p=0.031). However, similar performances were obtained across most of the datasets as sample sizes increased to around 400.

### 3.3 Classification Accuracies

We ran ExcCD, ExcD and IncD on 20 gene expression datasets (the datasets are listed in Table S1 in the Supplementary Materials). We found no statistically significant differences in accuracies according to a paired Wilcoxon signed rank test when using either the top 10, 20 or 30 genes (p>0.05; Table S2-3 in the Supplementary Materials). Table 1 summarizes the results of all of the experiments, where each cell contains the mean accuracy rank and corresponding standard deviation after training a support vector machine (SVM) with 20 random 50% train-50% test splits of the data.

|        | **ExcCD**   | **ExcD**    | **IncD**    |
|-------:|:-----------:|:-----------:|:-----------:|
| **Top 10** | 2.10(0.85) | 2.00(0.73) | 1.90(0.91) |
| **Top 20** | 1.80(0.89) | 2.25(0.85) | 1.95(0.69) |
| **Top 30** | 2.20(0.89) | 1.80(0.77) | 1.90(0.79) |

**Table 1.** Summary of SVM classification accuracies from real gene expression data. Variables were ranked using each of the three methods and the top 10, 20, or 30 variables were used to train a SVM. There were no significant differences in classification accuracies across the methods.

The training set was further split into a 40% training and a 10% validation set to tune the $C$ parameter of the SVM from {0.01, 0.1, 1, 10, 100} and the $\varepsilon$ parameter of K-CDM from {0.01, 0.001, 0.0001} by 5-fold cross-validation.

### 3.4 Transcription Factor-Gene Relationships

We used the ECOLI and YEAST datasets which contain 907 and 530 samples, respectively. We set each of the transcription factors in ECOLI and YEAST as a target and ran ExcCD, ExcD and IncD using 25, 50, 100, 200, 300, 400, and 500 samples on both datasets in order to identify the downstream target genes of each transcription factor. In addition, we ran 530 samples on ECOLI and 600, 700 and 907 samples on YEAST. The methods were thus run 138 and 107 times for ECOLI and YEAST respectively for each sample size. A summary of the results is presented in Figure 4. In YEAST, we found that ExcCD outperformed both ExcD and IncD according to a paired t-test at sample sizes of 500 (t(106)=-3.681, p<0.001; t(106)=-2.1299, p=0.036) and 530 (t(106)=-3.856, p<0.001; t(106)=-2.656, p=0.009). Similarly, ExcCD outperformed ExcD and IncD in ECOLI starting at a sample size of 600 (t(137)=-7.415, p<0.001; t(137)=-8.961, p<0.001).

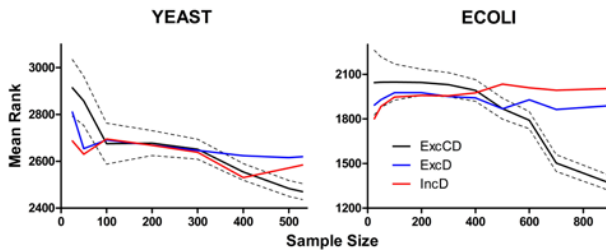

**Fig. 4.** Comparison of the ranks of the Markov blanket across ExcCD, ExcD and IncD. Solid lines represent mean rank and dotted lines represent the 95% confidence intervals. Note that ExcCD outperforms ExcD and IncD when sample sizes are above 400.

## 4 DISCUSSION

We developed an algorithm in this pauper which allowed us to compare the relative capabilities of DMs and CDMs in detecting $MB(Y)$. We also derived theoretical guarantees of ExcCD in the infinite sample case. In practice, we found that ExcCD outperforms ExcD and IncD in detecting $MB(Y)$ on most expert-designed and gene expression datasets when sample sizes are greater than several hundred. However, all of the methods had comparable accuracy when used as a feature selection method for an SVM suggesting that classification accuracy is a weak measure to assess accurate detection of $MB(Y)$ in real expression data.

The critical contribution of this paper is the ability to compare two schools of thought, one focusing on algorithms using DMs and the other on algorithms using CDMs. Many of these algorithms have been used interchangeably in bioinformatics without regard to the theoretical understanding of their outputs. While it may seem expected that dependence-based methods should outperform conditional dependence methods in the low sample size range, this has not been clearly elucidated in the past. For example, many papers describe selecting features from expression data by linear regression which is a measure of conditional dependence despite only having sample sizes in the few tens (Jin, et al., 2006; Muller, et al., 2008; Park and Nakai, 2011; Wu, 2006). In this case, a dependence measure such as the t-statistic would be more appropriate. We believe that the results presented in this paper suggest that the distinction between dependence and conditional dependence is important to help bioinformaticians better understand the causal interpretation of their models.

The need for larger sample sizes with ExcCD does not appear to be a result of the increased difficulty of assessing conditional dependence by the introduction of a third variable set $\{X \backslash X_s\}$. In this paper, both K-DM and K-CDM use two sets of variables, and the only difference between the two is a matrix inversion. Moreover, we assessed performance differences with a ranking criterion instead of an alpha level cut-off which allowed us to avoid the problem of obtaining an optimal cut-off value for each measure with a user-defined balance of Type I and Type II errors. We can thus conclude that using the residual errors as a feature ranking strategy in general makes $MB(Y)$ discovery difficult when samples are limited.

The ability of ExcCD to outperform HITON-MB, MMMB, and IAMB in discovering $MB(Y)$ with low sample sizes is a side-product of this work and can be better understood with several arguments. First, the latter three algorithms are divided into two stages, where the first stage identifies the direct causes and direct effects, and the second identifies the direct causes of the direct effects (Aliferis, et al., 2010). An error in the first stage can unfortunately increase the probability of error in the second. On the other hand, ExcCD is one stage by discovering all variables in $MB(Y)$ from the start. Next, ExcCD considers all possible conditioning sets while IAMB, HITON-MB and MMMB do not; in fact, conditioning on the entire dataset with HITON-MB and MMMB is not possible, since the time complexity is exponential with respect to the size of conditioning set (Statnikov, et al., 2013). Moreover, conditioning in IAMB is sample inefficient, since the algorithm only considers a test reliable when the number of instances in $X$ is at least 5 times the degrees of freedom in the test (Godo, 2005). On the other hand, ExcCD works by comparing the relative values of the CDMs between variables. As a result, reliability is established based on the relative differences between the variables instead of the degrees of freedom.

In conclusion, we developed a scalable algorithm to systematically assess the relative abilities of DMs and CDMs in detecting the $MB(Y)$ in gene expression data. We showed that methods using DMs are able to better discover variables in the $MB(Y)$ when sample sizes are low, but methods using CDMs perform the best when sample sizes are more than several hundred. No single strategy outperforms the other in expression data, so the choice of a particular method should be informed by the sample size at hand.

**ACKNOWLEDGEMENTS**

This research was funded by the National Library of Medicine grant T15 LM007059-24 to the University of Pittsburgh Biomedical Informatics Training Program and the National Institute of General Medical Sciences grant T32 GM008208 to the University of Pittsburgh Medical Scientist Training Program.

# Supplementary Materials

**Proposition.**

$$Y \perp\!\!\!\perp \{X \backslash MB(Y)\} | MB(Y) \Leftrightarrow Y \perp\!\!\!\perp X | MB(Y).$$

**Proof.**

First, consider the forward direction:

$$Y \perp\!\!\!\perp \{X \backslash MB(Y)\} | MB(Y) \Rightarrow P(Y|\{X \backslash MB(Y)\}, MB(Y)) = P(Y|X).$$

Since $X$ includes $MB(Y)$, we can state:

$$P(Y|X) = P(Y|\{X \backslash MB(Y)\}, MB(Y), MB(Y)) = P(Y|X, MB(Y)).$$

The equality $P(Y|X) = P(Y|X, MB(Y))$ implies that $Y \perp\!\!\!\perp X | MB(Y)$.

Second, the backward direction: $Y \perp\!\!\!\perp X | MB(Y) \Rightarrow P(Y|X, MB(Y)) = P(Y|X) = P(Y|\{X \backslash MB(Y)\}, MB(Y))$,

where the last equality implies that $Y \perp\!\!\!\perp \{X \backslash MB(Y)\} | MB(Y)$. □

| Dataset | Sample Size | Dimensions | Disease |
|---|---|---|---|
| 1. Chowdary, et al. (2006) | 104 | 22284 | Breast Cancer |
| 2. Golub, et al. (1999) | 72 | 7130 | Leukemia |
| 3. Subramanian, et al. (2005) | 50 | 10101 | N/A |
| 4. West, et al. (2001) | 49 | 7130 | Breast Cancer |
| 5. Su, et al. (2002) | 102 | 5566 | N/A |
| 6. Sorlie, et al. (2001) | 85 | 457 | Breast Cancer |
| 7. Nakayama, et al. (2007) | 105 | 22284 | Sarcoma |
| 8. Gravier, et al. (2010) | 168 | 2906 | Breast Cancer |
| 9. Khan, et al. (2001) | 63 | 2309 | Small Round Blue Cell Tumors |
| 10. Tian, et al. (2003) | 173 | 12626 | Myeloma |
| 11. Alon, et al. (1999) | 62 | 2001 | Colon Cancer |
| 12. Pomeroy, et al. (2002) | 60 | 7129 | Central Nervous System Embryonal Tumor |
| 13. Shipp, et al. (2002) | 77 | 7130 | Lymphoma |
| 14. Christensen, et al. (2009) | 217 | 1414 | N/A |
| 15. Gordon, et al. (2002) | 181 | 12534 | Lung Cancer |
| 16. Singh, et al. (2002) | 102 | 12601 | Prostate Cancer |
| 17. Yeoh, et al. (2002) | 248 | 12626 | Leukemia |
| 18. Burczynski, et al. (2006) | 127 | 22284 | Crohn's Disease |
| 19. Chiaretti, et al. (2004) | 128 | 12626 | Leukemia |
| 20. Sun, et al. (2006) | 180 | 54613 | Glioma |

**Table S1.** List of gene expression datasets used to assess differences in classification accuracy.

| Dataset | ExcCD | ExcD | IncD |
|---|---|---|---|
| 1. Chowdary | 97.02(2.29) | 95.10(3.50) | 95.58(2.80) |
| 2. Golub | 92.75(3.64) | 94.59(2.92) | 94.74(4.16) |
| 3. Subramanian | 66.13(8.95) | 67.28(5.65) | 64.05(9.72) |
| 4. West | 67.00(9.41) | 67.09(10.43) | 65.03(8.93) |
| 5. Su | 87.50(6.64) | 88.97(5.51) | 92.14(5.39) |
| 6. Sorlie | 73.00(6.36) | 73.36(4.31) | 73.73(5.61) |
| 7. Nakayama | 41.97(5.57) | 52.62(5.92) | 57.47(4.83) |
| 8. Gravier | 71.54(4.42) | 72.36(4.66) | 70.08(5.09) |
| 9. Khan | 89.28(6.80) | 90.14(7.71) | 92.56(5.05) |
| 10. Tian | 74.73(3.67) | 75.49(3.69) | 75.91(3.18) |
| 11. Alon | 80.48(5.18) | 76.45(7.91) | 75.81(10.07) |
| 12. Pomeroy | 61.27(5.36) | 62.55(10.90) | 59.37(6.91) |
| 13. Shipp | 85.23(4.98) | 80.77(5.98) | 80.80(5.75) |
| 14. Christensen | 97.87(1.24) | 98.30(1.81) | 98.30(1.50) |
| 15. Gordon | 98.73(0.82) | 98.56(1.30) | 98.45(1.21) |
| 16. Singh | 89.90(3.56) | 88.82(3.66) | 88.14(6.22) |
| 17. Yeoh | 79.47(5.11) | 82.97(3.35) | 83.99(3.41) |
| 18. Burczynski | 73.68(7.99) | 71.55(4.32) | 73.10(5.44) |
| 19. Chiaretti | 77.08(4.97) | 76.10(6.34) | 78.78(6.01) |
| 20. Sun | 62.06(3.65) | 61.83(3.23) | 64.48(4.95) |
| *Acc Rank* | 2.10(0.85) | 2.00(0.73) | 1.90(0.91) |

**Table S2.** Classification accuracies obtained when using the top 10 variables identified by each method.

| Dataset | ExcCD | ExcD | IncD |
|---|---|---|---|
| 1. Chowdary | 96.35(1.38) | 96.25(2.11) | 94.81(2.93) |
| 2. Golub | 94.29(3.23) | 96.52(2.72) | 95.94(2.95) |
| 3. Subramanian | 65.06(8.42) | 65.59(9.98) | 63.36(8.23) |
| 4. West | 67.21(7.59) | 61.30(8.12) | 64.07(11.43) |
| 5. Su | 92.87(4.07) | 92.29(4.53) | 95.96(3.56) |
| 6. Sorlie | 79.66(7.29) | 76.61(6.24) | 78.66(4.22) |
| 7. Nakayama | 53.78(6.64) | 56.96(6.32) | 60.39(5.05) |
| 8. Gravier | 72.79(4.94) | 72.37(4.52) | 72.21(5.84) |
| 9. Khan | 97.27(3.19) | 95.09(4.28) | 96.99(2.61) |
| 10. Tian | 77.30(2.61) | 76.24(4.14) | 76.83(3.43) |
| 11. Alon | 77.74(8.10) | 78.39(6.02) | 78.23(9.24) |
| 12. Pomeroy | 59.00(5.05) | 64.84(8.65) | 64.20(7.55) |
| 13. Shipp | 91.61(3.97) | 85.11(6.11) | 82.34(7.58) |
| 14. Christensen | 99.45(0.46) | 99.17(1.12) | 99.77(0.51) |
| 15. Gordon | 99.28(0.65) | 99.00(0.80) | 99.11(1.05) |
| 16. Singh | 89.80(3.95) | 91.08(2.95) | 90.39(3.24) |
| 17. Yeoh | 89.01(3.06) | 88.15(2.85) | 90.80(2.43) |
| 18. Burczynski | 78.54(8.30) | 75.73(6.03) | 78.46(3.50) |
| 19. Chiaretti | 83.36(4.71) | 80.45(5.15) | 82.12(4.29) |
| 20. Sun | 62.76(4.37) | 65.12(4.07) | 65.23(3.72) |
| *Acc Rank* | 1.80(0.89) | 2.25(0.85) | 1.95(0.69) |

**Table S3.** Classification accuracies obtained when using the top 20 variables identified by each method.

| Dataset | ExcCD | ExcD | IncD |
|---|---|---|---|
| 1. Chowdary | 97.79(1.43) | 95.48(2.59) | 95.77(2.97) |
| 2. Golub | 94.98(2.63) | 95.13(3.28) | 96.14(2.07) |
| 3. Subramanian | 68.55(6.90) | 68.00(6.55) | 63.60(9.60) |
| 4. West | 63.82(11.39) | 62.49(6.89) | 61.61(7.45) |
| 5. Su | 94.21(4.36) | 96.76(2.04) | 95.59(2.83) |
| 6. Sorlie | 76.79(6.73) | 78.97(5.45) | 80.67(4.92) |
| 7. Nakayama | 56.37(5.53) | 58.43(7.11) | 60.44(6.51) |
| 8. Gravier | 73.31(4.53) | 71.76(4.53) | 72.99(3.87) |
| 9. Khan | 97.10(3.76) | 98.57(2.18) | 97.47(2.42) |
| 10. Tian | 75.91(4.54) | 77.17(2.89) | 77.17(4.01) |
| 11. Alon | 78.06(7.66) | 81.77(5.26) | 79.68(4.92) |
| 12. Pomeroy | 59.49(6.90) | 61.20(8.00) | 57.41(7.99) |
| 13. Shipp | 88.45(5.23) | 87.24(6.70) | 86.05(6.64) |
| 14. Christensen | 99.77(0.41) | 99.91(0.28) | 100.00(0.00) |
| 15. Gordon | 98.62(1.29) | 99.23(0.72) | 98.67(1.17) |
| 16. Singh | 90.10(4.24) | 90.88(4.94) | 89.90(3.13) |
| 17. Yeoh | 94.91(1.74) | 92.15(2.17) | 93.29(1.57) |
| 18. Burczynski | 78.62(6.44) | 79.51(5.12) | 78.76(4.14) |
| 19. Chiaretti | 83.63(2.55) | 85.62(4.56) | 85.91(4.63) |
| 20. Sun | 63.61(3.54) | 63.50(3.06) | 65.70(4.24) |
| *Acc Rank* | 2.20(0.89) | 1.80(0.77) | 1.90(0.79) |

**Table S4.** Classification accuracies obtained when using the top 30 variables identified by each method.